\documentstyle[twocolumn,prb,aps]{revtex}
\begin{document}
\draft
\title{Fractional Quantum Hall States in Low-Zeeman-Energy Limit}
\author{X.G. Wu and J.K. Jain}
\address{Department of Physics, State University of New York
at Stony Brook, Stony Brook, New York 11794-3800}
\date{\today}
\maketitle
\begin{abstract}

We investigate the spectrum of interacting electrons at arbitrary
filling factors in the limit of vanishing Zeeman splitting.
The composite fermion theory successfully
explains the low-energy spectrum {\em provided the composite fermions
are treated as hard-core}.

\end{abstract}


\section{Introduction}

It is common to consider the limit of large magnetic field,
$B\rightarrow \infty$, in the theoretical study of fractional quantum
Hall effect (FQHE), since the phenomenon occurs at large magnetic
fields \cite {tsui}. This simplifies
the problem for two reasons. One, the Hilbert space is restricted
to the lowest Landau level (LL),
and consequently has a finite size. Two, the spin
degree of freedom is frozen, so the electrons can effectively be
taken to be spinless. However, the effective mass and the
effective g-factor in GaAs are such that the
Zeeman splitting is roughly 1/60 of the LL spacing \cite {halperin83}.
Therefore, while in the true $B\rightarrow\infty$ limit the Zeeman
splitting would maximally polarize the system, in
actual experiments it may actually be negligible. Indeed, it has been
found that at sufficiently low fields several incompressible FQHE
states are not maximally polarized \cite {expt}. Therefore, it
is meaningful to
consider the limit in which electrons are still restricted to the
lowest LL but the Zeeman splitting is zero, which we will call the
vanishing-Zeeman-splitting (VZS) limit. This paper reports our
investigation of this limit \cite {theory}. We restrict our discussion to
the lowest LL, i.e., to the filling factor range $\nu<2$. In fact, since
$2>\nu>1$ is related to $\nu<1$ by an exact particle-hole symmetry in
the lowest LL, it is sufficient to consider $\nu<1$.

The present study is a generalization of an earlier study of {\em spinless}
electrons \cite {dev}. There, it was found that the low-energy
Hilbert space of interacting electrons is well described at
{\em arbitrary filling factors} in terms of {\em non-interacting}
composite fermions (CF's). We find that the CF theory provides a
consistent picture of the low-energy states in the VZS
limit as well, except that here
we need to impose a hard-core condition on the CF's.

We start by giving a brief introduction to the CF theory of the
FQHE in Section II. Section III compares the numerical eigenstates with
the CF states for systems of six and eight electrons for a range of
filling factors. The paper is concluded in Section IV.

\section{Correlated Composite Fermion Basis}

The FQHE state is characterized by the formation of a new kind of
particle called composite fermion, \cite {cftheory} which is an
electron carrying two
(in general, an even number of) vortices of the wave function. In a
mean-field sense, a CF can be thought of as a bound state of an
electron and two flux quanta,
where a flux quantum is $\phi_{0}=hc/e$, since flux quanta
produce the same phase factors as vortices as the electrons wind
around each other. The strongly correlated
liquid of interacting electrons in the fractional quantum Hall state
is equivalent to a weakly interacing gas of CF's, and a good
qualitative as well as quantitative understanding is obtained
straightforwardly in terms of CF's. Because of the binding of the flux
quanta, the CF's see an effective  magnetic field \cite {cftheory}
\begin{equation}
B^*= B \mp 2 \phi_{0} \rho \;\;,
\label{effective}
\end{equation}
where $B$ is the external field, and $\rho$ is the electron (or CF)
density per unit area. The $-$ ($+$) sign corresponds to the case when
the flux bound to the CF's is in the same (opposite) direction as
$B^*$. This implies that the CF filling factor,
$\nu^*=\phi_{0}\rho/B^*$, is related to the electron filling factor,
$\nu=\phi_{0}\rho/B$, by
\begin{equation}
\nu=\frac{\nu^*}{2\nu^*\pm 1}\;\;.
\label{ff}
\end{equation}
The most fundamental consequence of the
formation of CF's is that, insofar as the low-energy dynamics is
concerned, the system of interacting electrons
behaves like a system of non-interacting fermionic
particles at a different filling
factor.  At the special electron filling factors $\nu=n/(2n\pm 1)$,
the CF's fill an integer number ($n$) of ``quasi-Landau levels" \cite
{footnote}.
This explains the origin of incompressibility in a partially filled
LL, and the observation of
FQHE at precisely these sequences of filling factors.
Transition from one FQHE plateau to another is expected to occur at
$\nu^*=n+1/2$, which is in excellent agreement with experiments \cite
{vjg}.  There is also convincing numerical evidence for the existence of
CF's. The exact low-energy spectra of interacting
electrons at $B$ look strikingly similar to those of
{\em non-interacting} particles at $B^*$ in finite system studies, and
the microscopic wave functions of the two systems are also very
closely related \cite {dev,wu}.  Recently,
Halperin, Lee and Read \cite {hlr} have proposed that the ``metallic"
state at $\nu=1/2$ can be understood as a fermi sea of CF's (i.e.,
CF's at zero magnetic field). A large number of recent
experiments provide further confirmation of the existence of CF's by
observing the composite fermi sea at $\nu=1/2$ \cite {du}.

We assume a large LL spacing throughout this work, which allows us to
restrict the Hilbert space to the lowest kinetic energy band.
For electrons with $\nu<2$, this implies
that the Hilbert space is restricted to the lowest LL. The electron
system maps to a CF system which involves, in general, several
quasi-LL's. In this case, the (restricted) Hilbert space consists
only of those states in which an integer number of
quasi-LL's is fully occupied, one quasi-LL is partially
occupied, and all higher quasi-LL's are unoccupied. The usefulness
of the CF theory lies in the fact that the size of the Hilbert
space at $\nu^*$ is much smaller than that at $\nu$,
resulting in a simplification of the problem.
More specifically, it provides a small correlated basis
for the low-energy eigenstates of interacting
electrons at $\nu$, called the CF basis,
and asserts that, insofar as the low-energy spectrum is
concerned, a good approximation of the exact eigenstates is
obtained by diagonalizing the Hamiltonian in this much smaller CF basis.
The CF basis states at $\nu^*$, which also provide the
correlated basis for interacting
electrons at $\nu$, are constructed as follows:

(i) First, determine $\nu^*$ from Eq.\ (\ref{ff}).
It is allowed to be negative.

(ii) The CF states at $\nu^*$ are related to the electron states at
$\nu^*$. In particular, non-interacting CF's at $\nu^*$
are related to non-interacting electrons at $\nu^*$.
Therefore, consider non-interacting electrons at $\nu^*$.
When $\nu^*$ is an even integer ($2n$), the ground state is unique,
and contains an integer number ($n$) of filled LL's.
In other cases, when $2n<\nu^*<2(n+1)$,
the ground state is highly degenerate, since all possible arrangements
of electrons in the partially filled $(n+1)$th LL have the same energy.
It is straightforward to write wave
functions for all these states. Let us denote these by $\chi_{\pm
\nu^*}^\alpha$.

(iii) The wave functions of non-interacting CF's at $\nu^*$ are now
obtained by simply multiplying these states by
$\Phi^{2}$, where
\begin{equation}
\Phi \equiv\prod_{j<k=1}^{N}(z_{j}-z_{k})\;.
\end{equation}
Here $z_{j}=x_{j}-iy_{j}$ denotes the position of the $j$th electron.

(iv) Finally, we project these product states onto the lowest LL.
Calling the
projection operator ${\cal P}$, we get the CF basis states:
\begin{equation}
\chi_{\nu}^\alpha= \Phi {\cal P} \Phi \chi_{\pm \nu^*}^\alpha\;\;.
\end{equation}
Note that we choose this projection as opposed to
the simple projection ${\cal P}\Phi^2
\chi_{\pm \nu^*}^\alpha$. This ensures that
$\chi_{\nu}^\alpha$ contains a factor of $\Phi$,
and satisfies the ``hard-core" property, i.e.,
has zero probability of having two electrons at the same point.
This builds good correlations in the presence of the repulsive
Coulomb interaction.

Multiplication by $\Phi^{2}$ attaches two vortices to each electron,
thus creating a CF.
Note that while non-interacting CF's at $\nu^*$ are related to
non-interacting electrons at $\nu^*$, they provide a correlated basis
for {\em interacting} electrons
at $\nu$. The CF theory thus maps the problem of interacting electrons
at $\nu$ to non-interacting electrons at $\nu^*$.

We employ the usual spherical geometry for our numerical calculations,
in which electrons move on the surface of a sphere, with the
magnetic field provided by a magnetic monopole
at the center\cite {haldane,yang}.
For a monopole of strength $q$, defined such that the
flux through the surface of the sphere is $2|q|\phi_{0}$, the
lowest LL single electron states have angular momentum
$l=|q|$, and the degeneracy of the lowest LL is $2(2|q|+1)$.
Thus the problem is that of $N$
interacting electrons in angular momentum $l=|q|$ shell.
The wave functions are a generalization of the spherical
harmonics, called the ``monopole harmonics" \cite {yang}.
The eigenstates of the many-body system have well-defined total
angular momentum, $L$, and total spin, $S$.

The CF theory is easily translated into the spherical geometry. Now
the low-energy states of interacting electrons at $q$ are related to
the low-energy  states of non-interacting electrons at $q^*$, given by
\begin{equation}
q=q^*+N-1\;\;,
\label{ffs}
\end{equation}
which is equivalent to Eq.~(\ref{ff}) in the limit of large $N$,
since, then, $\nu=N/2|q|$, and $\nu^*=N/2|q^*|$.
In the spherical geometry, $\Phi$ is the spatial part of the fully
polarized state at monopole flux strength $(N-1)/2$; it is completely
antisymmetric, and identical to
the wave function of the filled lowest LL of spinless electrons.
The CF basis states for interacting electrons at $q$ are then given by
\begin{equation}
\chi_{q}^\alpha= \Phi {\cal P} \Phi \chi_{\pm q^*}^\alpha\;\;.
\end{equation}
Since $\chi_{q}^\alpha$ are eigenstates of $L$ and $S$, we
choose $\chi_{q^*}^\alpha$ to be
eigenstates of $L$ and $S$. Multiplication by
$\Phi$ and projection onto the lowest LL do not change these quantum
numbers. Therefore, a state at $q^*$ with a given $L$ and $S$ produces
a state at $q$ with the same $L$ and $S$.
The states with different $L$-$S$ are automatically orthogonal,
so it is sufficient to diagonalize the Hamiltonian
separately in each $L$-$S$ subspace.

We have studied in the past a large number of filling factors for
6-10 electrons in the large-Zeeman-splitting limit, where $S$ takes
the largest possible value \cite{dev,wu}.
Our results convincingly showed the validity of the CF approach in
this limit.
We found that the low-energy states of interacting electrons at $q$
form a band, well separated from the other higher-energy states. The
number of states in this band, as well as their quantum numbers, match
perfectly with those of the CF basis states.  Furthermore, the actual
eigen-functions are very well approximated by the CF wave functions.

\section{Numerical Results}

In the present work, we set the Zeeman energy to zero.
We study a six electron system for $3 \leq q \leq 7$ and an eight
electron system for $5 \leq q \leq 6.5$.
Due to the symmetry of the problem, it is sufficient to
work in the sector where $L_{z}=0$ and $S_{z}=0$, where $L_{z}$ and
$S_{z}$ are the z-components of $L$ and $S$.
We restrict our discussion to this sector, with the
understanding that
when we talk about {\em one} state with a given $L$ and $S$ in this
sector, there are actually a total of $(2L+1)(2S+1)$ degenerate
states with the same energy.

The exact low-energy spectra for electrons interacting via the
Coulomb interaction are shown in Fig.1. The low-energy states are
expected to be related to the
low-energy states at $q^*$. The $L$-$S$ quantum numbers of the degenerate
ground states of non-interacting electrons at
$q^*$ are shown in Table I.

For six electrons at $q=4.0$ and 6.0, and for eight
electrons at $q=5.5$, the electron system maps to a CF system with the
filled lowest quasi-LL (i.e., $\nu^*=2$).
In these cases, there is only one CF state, with
quantum numbers 0-0, and it is expected to be
incompressible, i.e., separated from other states by a gap.
This is in agreement with the actual spectra of Fig.1.

In other cases, there is no satisfactory matching between the quantum
numbers of the CF's at $q^*$ and
those of the low-energy electron states at $q$.
We construct CF basis states according
to the above prescription. Many of these states are
annihilated upon projection on to the lowest LL; these are marked by
$A$ in Table I. Sometimes, there are two states at $q^*$
at a given $L$-$S$, which produce the same $L$-$S$ state at $q$; in
such a case, it is possible to construct two linear combinations of these
states so that one is annihilated.
The annihilation of some states brings the low-energy spectrum in agreement
with the CF theory for the six electron system at $q=3.0$ and 3.5,
and for the eight electron system at $q=5.0$.
However, in general, the situation is still unsatisfactory.

We now show that the CF theory explains the low-energy spectrum at
$q$, provided the CF's
are themselves taken to be interacting. Just as non-interacting CF's
are related to non-interacting electrons at $q^*$, interacting CF's
are related to interacting electrons at $q^*$.
For the present purpose, it is sufficient to incorporate
interactions only to the extent of distinguishing
hard-core states from other states, where, as mentioned earlier,
a hard-core state satisfies
the property that its wave function vanishes whenever {\em any} two
electrons coincide.

The Coulomb interaction is defined by its
pseudopotential parameters $V_{m}$ \cite
{haldane}, where  $V_{m}$ is the energy of two electrons in a state
of relative angular momentum $m$. Consider a model
interaction in which all $V_{m}$ are set to zero except $V_{0}$.
The hard-core states have zero energy for this model interaction
while the non-hard-core states have a finite positive energy.
Since $V_{0}$ is quite large
for the Coulomb interaction, we expect this model to be qualitatively
reasonable. We have found that the low-energy Coulomb states do indeed
satisfy the hard-core property to an
excellent approximation. Furthermore, these states form a well defined
band, the ``hard-core band", which is well separated from the
other non-hard-core states.
For example, for the six electron system at $q=3$ or 3.5, a low-energy band
is clearly visible. To make sure that the origin of this band lies in
the hard-core part of the Coulomb interaction, we
have constructed the true hard-core states
by diagonalizing the $V_{0}$ interaction. These have exactly the
same quantum numbers as the states in this band, and have very large
overlaps ($>0.99$) with these states.
For most values of $q$ considered here, the hard-core band is not
visible in Fig.1 since it is quite large,
and all of the states shown belong to this band.

Now let us consider hard-core CF's. This corresponds to taking the
electron states $\chi_{q^*}$ to be hard-core. ``Hard-core" is used
here in a slightly more general sense. We impose the hard-core
condition only on the electrons in the partially filled LL. This assumption
is valid when the hard-core interaction is small compared to the gap
between the quasi-LL's of the CF's.
Also, for more than half-filled LL, we impose the hard-core condition
on the holes rather than electrons. The
quantum numbers of the hard-core states at $q^*$ are marked by
an asterisk in Table I. These match quite well with the quantum
numbers of the low-energy states of interacting electrons at $q$;
the only exception is at $q=6$ for eight electrons,
where a 0-2 state, which is not a part of the CF basis,
has a slightly lower energy than the CF
state at 2-0. The overlaps of the CF states with the exact Coulomb
states are shown in Table II, and provide a more complete confirmation
of the CF theory \cite {repetition}.

Several comments are in order.

(i) In this work, no diagonalization of the Hamiltonian is necessary
in the hard-core CF basis. This is because, due to the small size of
the system, there is only one hard-core CF state at each $L$-$S$
studied here. Thus, the CF states do not contain any adjustable parameters.

(ii) Because of the factor $\Phi$, even the states of
non-interacting CF's satisfy the hard-core property for electrons
(provided they are not annihilated).
The hard-core interaction at $q^*$ (between CF's) corresponds
to a longer-range part of the inter-electron interaction at $q$.
When the CF's occupy more than one quasi-LL, the hard-core interaction
between the CF's is applicable only to the CF's in the
partially filled quasi-LL, which will translate into a rather complex
effective interaction between the electrons.

(iii) Annihilation of a large number of CF states may seem somewhat
mysterious. However, in most cases, it is explained  quite straightforwardly.
As indicated above, the CF basis states are hard-core by construction.
The projection operator {\em must} annihilate an unprojected
CF state when no hard-core state is available at $q$ at the
corresponding quantum numbers.  This is the reason for most of
the annihilations. Moreover, when there is only one hard-core $L$-$S$ state
at $q$, all unprojected CF states with these quantum numbers must
produce this state, as was found to be case at 0-0 and 1-1 for
the six electron system at $q=3$.
However, in some cases, e.g., for the 3-0
state of the six electron system at $q=3.5$, annihilation of the CF
state is non-trivial, since a 3-0 hard-core state {\em does} exist
here.

(iv) When there is only one hard-core $L$-$S$ state at $q$, the
projected CF state is identical to this state, and the large overlap
of this state with the Coulomb state tells us nothing more than that the
Coulomb state satisfies the hard-core property to a good
approximation. In other cases, there are several hard-core states at
$L$-$S$. For example, for the six electron system at $q=5$, there are eight
1-1 hard-core states. In such cases, the large overlaps provide a more
rigorous verification of the CF character of the low-energy states.

(v) The low-energy spectra at $q=N-1+q^*$ and $q=N-1-q^*$ look
strikingly similar, even though there is no {\em exact} symmetry relating
these two values of $q$. This is, of course, easily explained by the
CF theory.

(vi) Note that the hard-core property is satisfied by the CF's
automatically in the case of spinless electrons because
of the Pauli principle. Thus, the low-energy spectrum of interacting
electrons can be explained both in the large and small Zeeman energy
limits provided the CF's are taken to
be hard-core.

\section{Conclusion}

This work reports an extensive numerical study in the limit of
vanishing Zeeman splitting, and shows that the CF theory explains the
low-energy spectrum at arbitrary filling factors
provided a hard-core condition is imposed upon the
CF's.

This work was supported in part by the Office of Naval Research under
Grant no. N00014-93-1-0880, and by the National Science Foundation under
Grant No. DMR90-20637.

\begin{figure}
\caption{ This figure shows the low-energy spectra for several values of
$q$ for (a) six and (b) eight electrons. The spin quantum numbers of
some low-energy states are shown on the figure.}
\end{figure}

\vspace{1cm}

\begin{center}

\begin{table}

\begin{tabular}{|c|c|l|} \hline
$q$  &  $q^*$  &  $L$-$S$ \\ \hline
     &         & 2-2$^*$, 0-0$^*$, 1-1$^*$, 2-1(A), 3-0(A), 4-1(A), \\
3.0  & -2.0    & 5-1(A), 6-0(A), 0-0(A), 1-1(A), 2-0(A), 2-0(A), \\
     &         & 3-1(A), 3-1(A), 4-0(A), 4-0(A) \\ \hline
3.5  & -1.5    & 0-1$^*$, 1-0$^*$, 2-1$^*$, 3-0(A)  \\ \hline
4.0  & -1.0    & 0-0$^*$ \\ \hline
4.5  & -0.5    & 0-1$^*$, 2-1$^*$, 1-0, 3-0 \\ \hline
5.0  & +0.0    & 1-1$^*$, 0-0, 2-0 \\ \hline
5.5  & +0.5    & 0-1$^*$, 2-1$^*$, 1-0, 3-0 \\ \hline
6.0  & +1.0    & 0-0$^*$ \\ \hline
6.5  & +1.5    & 0-1$^*$, 1-0$^*$, 2-1$^*$, 3-0  \\ \hline
     &         & 2-2$^*$, 0-0$^*$, 1-1$^*$, 2-1, 3-0, 4-1, \\
7.0  & +2.0    & 5-1, 6-0, 0-0, 1-1, 2-0, 2-0, \\
     &         & 3-1, 3-1, 4-0, 4-0 \\ \hline
\end{tabular}

\vspace{0.5cm}
TABLE I(a)
\vspace{1.5cm}

\begin{tabular}{|c|c|l|} \hline
$q$  &  $q^*$  &  $L$-$S$ \\ \hline
5.0  & -2.0    & 0-0$^*$, 2-0$^*$, 3-1$^*$, 1-1$^*$, 4-0(A) \\ \hline
5.5  & -1.5    & 0-0$^*$ \\ \hline
6.0  & -1.0    & 1-1$^*$, 3-1$^*$, 0-0$^*$, 2-0$^*$, 4-0 \\ \hline
6.5  & -0.5    & 0-2$^*$, 3-1, 2-1, 1-1, 4-0, 2-0, 0-0, 2-0 \\ \hline
\end{tabular}

\vspace{0.5cm}
TABLE I(b)
\vspace{1.5cm}

\caption{ This table shows the quantum numbers of
all states with the lowest kinetic energy
at $q^*$ for (a) six and (b) eight
particles. The states satisfying the hard-core property are marked by
asterisk. The states marked by (A) are annihilated upon the CF
transformation, and do not produce any CF state at $q$.}

\end{table}

\vspace{1.5cm}
\begin{table}

\begin{tabular}{|c|c|c|c|c|c|} \hline
$q$/$L$-$S$ & 3.0/0-0 & 3.0/1-1 & 3.0/2-2 & 3.5/0-1 & 3.5/1-0 \\
\hline
overlap & 0.9991 & 0.9993 & 0.9988 & 0.9959 & 0.9978 \\ \hline \hline
$q$/$L$-$S$ & 3.5/2-1& 4.0/0-0 & 4.5/0-1 & 4.5/2-1 & 5.0/1-1 \\ \hline
overlap & 0.9970 & 0.9990 & 0.9956 & 0.9928 &0.9879 \\ \hline \hline
$q$/$L$-$S$ & 5.5/0-1 & 5.5/2-1 & 6.0/0-0 & 6.5/0-1 & 6.5/1-0 \\
\hline
overlap  & 0.9324 & 0.9768 & 0.9812 & 0.9696 & 0.9832 \\ \hline \hline
$q$/$L$-$S$ &6.5/2-1 & 7.0/0-0 & 7.0/1-1 & 7.0/2.2 & \\ \hline
overlap & 0.9848 & 0.9937 & 0.9934 & 0.9881& \\ \hline
\end{tabular}

\vspace{0.5cm}
TABLE II(a)
\vspace{1.5cm}

\begin{tabular}{|c|c|c|c|c|c|} \hline
$q$/$L$-$S$ & 5.0/0-0 & 5.0/2-0 & 5.0/3-1& 5.0/1-1 & 5.5/0-0 \\ \hline
overlap & 0.9930 & 0.9959 & 0.9940 & 0.9919 & 0.9980 \\ \hline \hline
$q$/$L$-$S$ &6.0/1-1 & 6.0/3-1& 6.0/0-0 & 6.0/2-0 & 6.5/0-2  \\ \hline
overlap &0.9807 & 0.9906 & 0.9829 & 0.9466 & 0.9918  \\ \hline
\end{tabular}

\vspace{0.5cm}
TABLE II(b)
\vspace{1.5cm}

\caption{This table gives the overlaps between the hard-core CF
states and the corresponding exact Coulomb states for (a) six and (b) eight
electrons.}

\end{table}

\end{center}


\begin{thebibliography}{99}

\bibitem{tsui} D.C. Tsui, H.L. Stormer, and A.C. Gossard, Phys. Rev.
Lett. {\bf 48}, 1559 (1982).

\bibitem{halperin83} B.I. Halperin, Helv. Phys. Acta {\bf 56}, 75 (1983).

\bibitem{expt} R.G. Clark, S.R. Haynes, A.M. Suckling, J.R. Mallett, P.A.
Wright, J.J. Harris, and C.T. Foxon, Phys. Rev. Lett. {\bf 62}, 1536
(1989); J.P. Eisenstein, H.L. Stormer, L.N. Pfeiffer, and K.W. West,
Phys. Rev. Lett. {\bf 62}, 1540 (1989);
J.P. Eisenstein, H.L. Stormer, L.N. Pfeiffer, and K.W. West,
Phys. Rev. B{\bf 41}, 7910 (1990);
L. W. Engel, S.W. Hwang, T. Sajoto, D.C. Tsui, and M. Shayegan,
Phys. Rev. B {\bf 45}, 3418 (1992).

\bibitem{theory} For earlier calculations in the VZS limit, see:
F.C. Zhang and T. Chakraborty, Phys. Rev. B {\bf 30}, 7320
(1984); T. Chakraborty and F.C. Zhang, Phys. Rev. B {\bf 29}, 7032
(1984); E.H. Rezayi, Phys. Rev. B {\bf 36}, 5454 (1987);
X.C. Xie, Y. Guo, and F.C. Zhang, Phys. Rev. B {\bf 40}, 3487
(1989); T. Chakraborty, Surf. Sci, {\bf 229}, 16 (1990);
X.C. Xie and F.C. Zhang, Mod. Phys. Lett. B {\bf }, 471
(1991);
P.A. Maksym, J. Phys. Condens. Matter {\bf 1}, 6299 (1989); T.
Chakraborty and P. Pietilainen, Phys. Rev. B {\bf 41}, 10862 (1990).

\bibitem{dev} G. Dev and J.K. Jain, Phys. Rev. Lett {\bf 69}, 2843 (1992).

\bibitem{cftheory} J.K. Jain, Phys. Rev. Lett. {\bf 63}, 199 (1989);
Phys. Rev. B {\bf 41}, 7653 (1990); Adv. Phys. {\bf 41}, 105 (1992).

\bibitem{footnote} We use the term ``quasi-LL" for the energy levels
of composite fermions to emphasize that these are different
from the {\em real} LL's of electrons.
While the real LL's of electrons occur as a result of the
kinetic energy quantization, quasi-LL's of CF's occur
as a result of an ``effective" quantization of the
interaction energy. Of course, CF's can occupy several quasi-LL's even
when the electrons are completely confined to their lowest {\em real} LL

\bibitem{vjg} V.J. Goldman, J.K. Jain, and M. Shayegan,
Phys. Rev. Lett. {\bf 65}, 907 (1990); Mod. Phys. Lett. {\bf 5}, 479
(1991).

\bibitem{wu} X.G. Wu, G. Dev, and J.K. Jain, Phys. Rev. Lett. {\bf
71}, 153 (1993).

\bibitem{hlr} B.I. Halperin, P.A. Lee, and N. Read, Phys. Rev. B {\bf
47}, 7312 (1993).

\bibitem{du} R.R. Du, H.L. Stormer, D.C. Tsui, L.N. Pfeiffer, and K.W.
West, Phys. Rev. Lett. {\bf 70}, 2944 (1993); R.L. Willett, R.R. Ruel,
K.W. West, and L.N. Pfeiffer, preprint;
W. Kang, H.L. Stormer, L.N. Pfeiffer, K.W. Baldwin, and K.W.
West, preprint; V.J. Goldman, Bo Su, and J.K. Jain, preprint.


\bibitem{haldane} F.D.M. Haldane, Phys. Rev. Lett. {\bf 51}, 605 (1983).

\bibitem{yang} T.T.  Wu and C.N. Yang, Nucl. Phys. B {\bf 107}, 365
(1976); Phys. Rev. D {\bf 16}, 1018 (1977).

\bibitem{repetition} The results for the incompressible states have
been published earlier \cite {wu},
and are repeated here for completeness.
Also, the spin-singlet CF state at $q=6$ for $N=6$ is identical to
one of the states proposed by Halperin \cite {halperin83}.


\end{thebibliography}
\end{document}